\documentclass[12pt]{article}
\usepackage[dvips]{color}
\usepackage{amsmath}
\usepackage{graphicx}
\usepackage{subfigure}
\usepackage{epsfig}
\usepackage{amsmath}
\usepackage{cite}
\usepackage{color}
\usepackage{subfigure}

\begin{document}
\begin{center}
\Large{\bf Thermodynamic Topology of Quantum Corrected\\ AdS-Reissner-Nordstrom Black Holes in Kiselev Spacetime}\\
\small \vspace{1cm} {\bf Jafar Sadeghi$^{\star}$\footnote {Email:~~~pouriya@ipm.ir}}, \quad
{\bf Saeed Noori Gashti$^{\dag}$\footnote {Email:~~~saeed.noorigashti@stu.umz.ac.ir}}, \quad
{\bf M. R. Alipour$^{\star}$\footnote {Email:~~~mr.alipour@stu.umz.ac.ir}}, \quad
{\bf M. A. S. Afshar$^{\star}$\footnote {Email:~~~m.a.s.afshar@gmail.com}}, \quad
 \\
\vspace{0.5cm}$^{\star}${Department of Physics, Faculty of Basic
Sciences,
University of Mazandaran\\
P. O. Box 47416-95447, Babolsar, Iran}\\
\vspace{0.5cm}$^{\dag}${School of Physics, Damghan University, P. O. Box 3671641167, Damghan, Iran}\\
\small \vspace{1cm}
\end{center}
\begin{abstract}
In this paper, we delve into the intricate thermodynamic topology of quantum-corrected Anti-de Sitter-
Reissner-Nordstrm (AdS-RN) black hole within the framework of Kiselev spacetime. By employing the generalized off-shell Helmholtz free energy approach, we meticulously compute the thermodynamic topology of these selected black holes. Furthermore, we establish their topological classifications. Our findings reveal that quantum correction terms influence the topological charges of black holes in Kiselev spacetime, leading to novel insights into topological classifications. Our research findings elucidate that, in contrast to the scenario in which $\omega=0$ and $a=0.7$ with total topological charge $W=0$ and $\omega=-4/3$ with total topological charge $W=-1$, in other cases, the total topological charge for the black hole under consideration predominantly stabilizes at +1. This stabilization occurs with the significant influence of the parameters a, c, and $\omega$ on the number of topological charges. Specifically, when $\omega$  assumes the values of $\omega=-1/3$, $\omega=-2/3$ , $\omega=-1$,  the total topological charge consistently be to W = +1.
\end{abstract}
\section{Introduction}
Black hole thermodynamics is a field of physics that explores the connection between the laws of thermodynamics
and the properties of black holes. Black holes have thermodynamic quantities such as entropy and temperature,
which are related to classical attributes such as horizon area and surface gravity. Black hole thermodynamics
involves integrating general relativity, quantum mechanics, and thermodynamics into a comprehensive description of black holes. The motivation for studying black hole thermodynamics comes from several sources, such
as The analogy between the laws of black hole mechanics and the laws of thermodynamics. The discovery of
Hawking radiation shows that black holes emit thermal radiation and have a finite temperature. The
holographic principle, suggests that the information content of a region of space is proportional to its boundary
area rather than its volume. The AdS/CFT correspondence, which relates a gravitational theory in anti-de
Sitter (AdS) space to a conformal field theory (CFT) on its boundary. in black hole thermodynamic studies,
some concepts are very important including the critical points and phase transition. A critical point is a point
in the phase diagram of a system where the distinction between different phases disappears. For black holes in
AdS space, there is a critical point where the small and large black holes have the same temperature and free
energy. At this point, the heat capacity of the black holes diverges, indicating a second-order phase transition.
The Hawking-Page phase transition is a phase transition between AdS black holes with radiation and thermal
AdS. Hawking and Page showed that although AdS black holes can be in stable thermal equilibrium with radiation, they are not the preferred state below a certain critical temperature. At this temperature, there will be a first-order phase transition where below $T_c$, thermal AdS will become the dominant contribution to the
partition function. In the context of the AdS/CFT correspondence, this phase transition can be interpreted as
the confinement-deconfinement phase transition of the strongly coupled field theory on the boundary \cite{1,2,3,4,5,6,7,8,9,10,11,12,13,14,15,16,17,18,19,20}.\\\\
The Thermodynamics topology of black holes is a topic that studies the topological properties of black hole thermodynamics, such as the critical points, phase transitions, and stability. Recently, new studies have been done on critical points and phase transitions in the thermodynamics of black holes from a topological perspective. The topological classification of thermodynamics of black holes is a method to study the phase transitions and critical points of black holes using topological tools such as winding numbers, Brouwer degrees, and Duan's $\phi$-mapping
method. These tools can reveal the local and global topological properties of the black hole spacetime and
its thermodynamic state space. Different types of black holes, such as Kerr-AdS, Kerr-Newman-AdS, Euler-
Heisenberg, Young Mills, and Banados-Teitelboim-Zanelli can have different topological classes depending on
their parameters and dimensions. The topological classification can help us understand the intrinsic properties of black holes under smooth deformations and the effects of cosmological constants and other features on their thermodynamics. Researchers have done some work by using the entropy-temperature (Duan's topological current $\phi$-mapping theory) \cite{21,22,23,24,25} and generalized Helmholtz free energy \cite{45}. Also for further study, you can see\cite{26,27,28,29,30,31,32,33,34,35,36,37,38,39,40,41,42,43}. Also, recently, topology structures and photon spheres have been used to determine the range of parameters for black holes\cite{44,44'}.\\\\
Quantum corrections play a pivotal role in addressing the singularity issue inherent in classical general relativity. The traditional Schwarzschild solution presents a singularity at r = 0, where spacetime curvature
escalates to infinity. D.I. Kazakov and S.N. Solodukhin have examined spherically symmetric quantum
fluctuations in both the metric and matter fields. Their work leads to an effective two-dimensional dilaton gravity
model\cite{46}. These quantum adjustments modify the Schwarzschild solution, effectively replacing the classical
singularity at r = 0 with a quantum-corrected zone. This area has a minimum radius, $r_{min}$, on the order of
the Planck length, $r_{Pl}$, ensuring the scalar curvature remains finite. The significance of this modification lies in its suggestion of a singularity-free, regular spacetime. The spacetime is composed of two asymptotically regions joined at a hypersurface of constant radius. This indicates that quantum effects could smooth out the singularities that classical general relativity predicts, thus enriching our comprehension of black hole
physics and the quantum-level structure of spacetime. Consequently, the investigation into quantum corrections has attracted considerable interest among researchers, sparking studies in various domains. These include
the criticality and efficiency of black holes, the thermodynamics of a quantum-corrected Schwarzschild black
hole in the presence of quintessence, accretion processes onto a Schwarzschild black hole within a quintessence
environment, and explorations of quasinormal modes, scattering, shadows, and the Joule-Thomson effect\cite{47,48,49,50,51}.\\\\
Motivated by the above concepts, we aim to examine the thermodynamic topology and the topological classes
of the quantum corrected AdS-(RN) Black Holes in Kiselev spacetime in this paper. We will also compare our results with other works in the literature. The paper is organized as follows. In section 2, we review the thermodynamics of the quantum-corrected AdS-(RN) Black Holes in Kiselev spacetime. In section 3, we introduce the basic notions of thermodynamic topology. In section 4, we analyze the thermodynamic topology of the quantum corrected AdS-(RN) Black Holes in Kiselev spacetime from the perspectives of generalized Helmholtz free energy. Finally, we summarize our findings and conclusions in section 5.
\section{The Model}
In this section, we will thoroughly examine the quantum-corrected AdS-RN black hole, which is encapsulated
within Kiselev spacetime. Our objective is to perform an exhaustive analysis of its attributes, particularly the
thermodynamic topology from the perspectives of generalized Helmholtz free energy, which is of paramount
importance to our study. We will explore the spacetime metric of the quantum-corrected charged AdS black
hole, encircled by a Kiselev spacetime. This metric is distinguished by its spherical symmetry and is articulated
as follows \cite{51},
\begin{equation}\label{eq1}
ds^2=f(r)dt^2-f(r)^{-1}dr^2-r^2d\Omega^2,
\end{equation}
So with respect to\cite{51},
\begin{equation}\label{eq2}
\begin{split}
f(r)=-\frac{2M}{r}+\frac{\sqrt{r^2-a^2}}{r}+\frac{r^2}{\ell^2}-\frac{c}{r^{3\omega+1}}+\frac{Q^2}{r^2},
\end{split}
\end{equation}
Here, we note that \( r > a \) is used to prevent the formation of imaginary structures.  In other words, we consider the quantum fluctuation effects when $r$ is approximately larger than $a$. In our research, we aim to clarify the parameters that define the black hole in question. The parameter M denotes
the black hole's mass, while the parameter $a$  to the quantum corrections affect the black hole's characteristics. The symbol $\ell$ represents the length scale relevant to the asymptotically AdS (Anti-de Sitter) spacetime. The parameter $c$ is associated with the cosmological fluid encircling the black hole, and $Q$ indicates the black hole's electric charge. To initiate our discussion, comprehending the rationale for choosing the specific metric and analyzing the origin of each term is crucial. M. Visser has proposed that the Kiselev black hole model can be expanded to include a spacetime with N components. This extension is marked by a linear correlation between
energy and pressure for each component, as detailed in the cited literature\cite{52,53}. In our analysis, we consider various values for $\omega$, such as $\omega=-1/3$, $\omega=-2/3$, $\omega=-1$ and $\omega=-4/3$. The parameter a is intricately connected to changes in the black hole's mass due to quantum corrections. The underlying theory for this parameter is extensively discussed in\cite{54,55}. As an independent variable, a possesses the distinctive feature that, when set to zero, the metric reverts to the familiar AdS-Reissner-Nordstrm metric, now shrouded by a cosmic fluid. Theoretically, a can assume any value as long as it is smaller than the event horizon's radius, aligning with the notion that a constitutes a minor modification to the conventional black hole metric,
\begin{equation}\label{eq3}
\begin{split}
dM=TdS+VdP+\phi dQ+\mathcal{C}dc+\mathcal{A}da.
\end{split}
\end{equation}
To determine the entropy of the quantum-corrected Schwarzschild black hole located in Kiselev spacetime, it is
noted in\cite{6} that this entropy is consistent with the Hawking-Bekenstein entropy formula. As a result,
\begin{equation}\label{eq4}
\begin{split}
S=\frac{A}{4}=\pi r_+^2.
\end{split}
\end{equation}
The formula for pressure is given by $P = 3/8\pi\ell^2$ \cite{56}. The mass and Hawking temperature of
the quantum-corrected AdS-RN black hole, enveloped by Kiselev spacetime, is calculated as follows,
\begin{equation}\label{eq5}
\begin{split}
M=\frac{1}{2\sqrt{\pi}}\bigg(\sqrt{S-\pi a^2}-c\pi^{\frac{3\omega+1}{2}}S^{-\frac{3\omega}{2}}+\frac{8PS^{3/2}}{3}+\pi Q^2S^{-1/2}\bigg),
\end{split}
\end{equation}
and
\begin{equation}\label{eq6}
T_{H}=\big(\frac{\partial M}{\partial S}\big)_{P,Q}=\frac{1}{4\sqrt{\pi}}\bigg(\frac{1}{\sqrt{S-\pi a^2}}+8P\sqrt{S}+3\frac{c\omega}{\sqrt{\pi}}\big(\frac{\pi}{S}\big)^{\frac{3\omega}{2}+1}-\frac{\pi Q^2}{S^{3/2}}\bigg).
\end{equation}
The first law of thermodynamics robustly accommodates variations in the black hole's defining parameters, or
hair, which include the black hole's area, the cosmological constant, the electric charge, the quintessence parameter, and the quantum correction parameter. For a comprehensive discussion on integrating the quintessence
parameter as a thermodynamic variable. Our study further develops this framework to include the variability
of the quantum correction parameter.
\section{Thermodynamic topology}
Thermodynamic topology is a method that integrates topology into the study of black hole thermodynamics.
This approach involves assigning topological numbers to each zero point in the phase diagram. The topological
number is determined as the residue of the generalized Helmholtz free energy at the critical point. This unique
number can unveil new features and classifications of black hole thermodynamics that conventional methods
fail to capture. For instance, it can differentiate between conventional and novel critical points, each bearing
distinct implications for the first-order phase transition. The Helmholtz off-shell free energy is an extension of
the Helmholtz free energy that accommodates non-equilibrium states of a system. The conventional Helmholtz
free energy is the internal energy of the system subtracted by the product of the temperature and the entropy
of the system, representing the amount of useful work obtainable from a closed thermodynamic system at a
constant temperature and volume. In contrast, the Helmholtz off-shell free energy is the Legendre transform of
the internal energy concerning the entropy, which can be expressed as\cite{45,57,58,59,60,61}
\begin{equation}\label{eq7}
F(S,V,Y)=U(S,V,Y)-TS.
\end{equation}
In the equation, S represents the entropy, V denotes the volume, Y encompasses other extensive variables,
T stands for the temperature, and U signifies the internal energy. The Helmholtz off-shell free energy is
instrumental in analyzing phase transitions and critical points of black holes within various frameworks, such as
anti-de Sitter (AdS) or de Sitter (dS) spaces, and can accommodate scenarios with or without electric charge,
including nonlinear electromagnetic fields. To elucidate the thermodynamic properties of black holes, we employ
diverse quantities. For instance, mass and temperature can serve as two variables to articulate the generalized
free energy. Drawing upon the mass-energy equivalence in black hole physics, we can recast our generalized free
energy function into a conventional thermodynamic function, as follows\cite{45,57,58,59,60,61},
\begin{equation}\label{eq8}
\mathcal{F}=M-\frac{S}{\tau},
\end{equation}
where $\tau$ represents the Euclidean time period and T, which is the inverse of $\tau$, denotes the temperature of the ensemble. The vector $\phi$ is defined as follows\cite{45,57,58,59,60,61},
\begin{equation}\label{eq9}
\phi=(\phi^{r_H},\phi^\Theta)=\big(\frac{\partial\mathcal{F}}{\partial r_{H}},-\cot\Theta\csc\Theta\big),
\end{equation}
Here, $r_H$ represents the radius of the event horizon, and the parameter $0 \leq \Theta \leq \pi$ is introduced due to the axis limit\cite{45}. It should be noted that the vector field $\phi$ points outward at the points where $\Theta = 0$ and $\pi$, due to the divergence of the component $\phi$ at these points. This is analyzed using Duan’s $\phi$-mapping topological current. The vector $\phi_\Theta$ has infinite magnitude and points away from the origin when $\Theta$ is either 0 or $\pi$. The variables $r_H$ and $\Theta$ can take any values from 0 to infinity and from 0 to $\pi$, respectively. Additionally, we can rewrite the vector as $\phi = ||\phi|| e^{i\Theta}$, where $||\phi|| = \sqrt{\phi_a \phi_a}$, or $\phi = \phi^{r_H} + i\phi$. Topology can be applied to black hole thermodynamics to classify the critical points and phase transitions of black hole systems according to their topological charges and numbers. A critical point is a point in the phase diagram where the system changes state or behavior, such as a phase transition or a stability change. A topological charge is a quantity that measures the winding number of a vector field around a critical point. A winding number is an integer that counts how many times a vector field wraps around a point in a plane. A topological number is the sum of all the topological charges in a system, which reflects the global topological nature of the system. One way to perform the topological analysis of black hole thermodynamics is to use Duan's topological current $\phi$-mapping theory. This theory maps the thermodynamic variables to a vector field $\phi$ in a two-dimensional plane, and defines a topological current $j^{\mu}$ as\cite{45,57,58,59,60,61},
\begin{equation}\label{eq10}
j^{\mu}=\frac{1}{2\pi}\varepsilon^{\mu\nu\rho}\varepsilon_{ab}\partial_{\nu}n^{a}\partial_{\rho}n^{b},\hspace{1cm}\mu,\nu,\rho=0,1,2,,
\end{equation}
where $n = (n^1, n^2)$, we have $n^1 = \frac{\phi^r}{\|\phi\|}$ and $n^2 = \frac{\phi^\theta}{\|\phi\|}$. The topological current is nonzero only at the zero points of $\phi$, which correspond to the critical points of the thermodynamic system. The topological charge $Q$ at each critical point is given by\cite{45,57,58,59,60,61},
\begin{equation}\label{eq11}
Q_t=\int_\Sigma\Sigma_{i=1}^{n}\beta_{i}\eta_{i}\delta^2(\overrightarrow{x}-\overrightarrow{z})_{i}=\Sigma_{i=1}^{n}\beta_{i}\eta_{i}=\Sigma_{i=1}^{n}\widetilde{\omega}_{i},
\end{equation}
where $\beta_i$ is the positive Hopf index, which counts the loops of the vector $\phi_a$ in the $\phi$ space when $x^\mu$ is near the zero point $z_i$. Meanwhile, $\eta_i = \text{sign}(j_0(\phi/x)_{z_i}) = \pm 1$. The quantity $\widetilde{\omega}_{i}$ represents the winding number for the $i$-th zero point of $\phi$. The winding number can be calculated by using the following formula\cite{45,57,58,59,60,61},
\begin{equation}\label{eq12}
W=\frac{1}{2\pi}\int_{c_{i}}d\Omega,
\end{equation}
Then the total charge will be,
\begin{equation}\label{eq13}
Q=\sum_{i}\widetilde{\omega}_{i}.
\end{equation}
Using this method, one can classify different types of black holes according to their topological charges and
numbers. The generalized Helmholtz free energy method is another way to perform the topological analysis of
black hole thermodynamics. It is based on the generalized off-shell Helmholtz free energy, which is a function
of the thermodynamic variables that is valid for any state of the system, not only for the equilibrium states.
Using this method, one can plot the generalized off-shell Helmholtz free energy as a function of V for a fixed
T, and identify the critical points and phase transitions of the black hole system by looking at the shape and
features of the curve.
\section{Discussion and Result}
Topological techniques are instrumental in analyzing the critical points and phase transitions within black
hole thermodynamics. This is achieved by attributing a topological invariant to each critical point on the
phase diagram. A topological invariant is a numerical value that delineates the nature and sequence of the
phase transition occurring at a critical juncture. This invariance can be determined through the application
of the residue theorem or by employing Duan's topological current theory. In this section, our objective is to
scrutinize the thermodynamic topology of the aforementioned black holes. We will elucidate the specifics of our
computational approach subsequently. Consequently, in accordance with the preceding equation, the Helmholtz
free energy for this black hole is deduced as follows,
\begin{equation}\label{eq14}
\mathcal{F}=\frac{4 r^{-3 w-1} \left(\frac{3}{8} \pi  P \sqrt{r^2-a^2} r^{3 w+1}-\frac{3}{8} \pi  c P r+\frac{3}{8} \pi  P Q^2 r^{3 w}+r^{3 w+4}\right)}{3 \pi  P}-\frac{\pi  r^2}{\tau }.
\end{equation}
Two vector fields, $\phi^{rh}$ and $\phi^\theta$, are calculated in accordance with the concepts previously discussed, as follows,
\begin{equation}\label{eq15}
\phi^{rh}=\frac{r}{2 \sqrt{r^2-a^2}}+\frac{3}{2} c \omega r^{-3 \omega-1}+\frac{4 r^2}{\pi  p}-\frac{Q^2}{2 r^2}-\frac{2 \pi  r}{\tau },
\end{equation}
and
\begin{equation}\label{eq16}
\phi^\theta=-\frac{\cot\theta}{\sin\theta},
\end{equation}
Also, we obtain the $\tau$ as follows,
\begin{equation}\label{eq17}
\tau=\frac{4 \pi ^2 P \sqrt{r^2-a^2} r^{3 \omega+3}}{3 \pi  c P r \omega \sqrt{r^2-a^2}+\pi  (-P) Q^2 \sqrt{r^2-a^2} r^{3 \omega}+8 \sqrt{r^2-a^2} r^{3 \omega+4}+\pi  P r^{3 \omega+3}},
\end{equation}
In our study, we delve into the thermodynamic topology associated with a quantum-corrected charged AdS black hole enveloped by Kiselev spacetime. The illustrations are bifurcated, showcasing the normalized field lines on the right. The illustrations reveal a singular zero point in figures (1b, 1f, 2b, 2d, 4b, 4d, 5b), indicative of the one topological charge determined by the free parameters mentioned in the study. This charge correlates with the winding number and resides within the blue contour loops at coordinates $(r,\theta)$. The sequence of the illustrations is dictated by the parameter $\omega$; for instance, in figure (1), $\omega$ is set to $-\frac{1}{3}$, and for figures 2 through 5, $\omega$ assumes the values $-\frac{2}{3}$, $-\frac{4}{3}$, -1, 0 respectively. To construct these contours, we selected free parameters $(Q= 1; a = 0.4, 0.7; c = 0.4, 0.7)$. The findings from figures (1b, 1f, 2b, 2d, 4b, 4d, 5b) reveal that the distinctive feature of a positive topological charge of unity is the zero point enclosed within the contour. Our discourse explores the black hole stability by scrutinizing the winding numbers alongside the specific heat capacity. The affirmative winding numbers infer the thermodynamic stability of the on-shell black hole, which is further substantiated by the specific heat capacity calculations. Given the solitary on-shell black hole, its topological number aligns with the winding number, amounting to 1. This denotes the presence of a single stable on-shell black hole, equating to a topological number that mirrors a positive winding number across all BH configurations $(W = \widetilde{\omega} = +1)$. Conversely, figures (3b,3c), delineated for the free parameters (Q = 1; a = 0.4; c = 0.7), portrays three topological charges $(\widetilde{\omega} = -1,+1,-1)$, culminating in a total topological charge of W = -1. This is a departure from the preceding figures and transpires when $\omega=-4/3$ for the specified black hole. Moreover, figure (1d), crafted for free parameters Q = 1; a = 0.7; c = 0.4, and figures (5d,5e), for free parameters (Q = 1; a = 0.7; c = 0), exhibit four topological charges ($\widetilde{\omega} = -1,+1, -1,+1$), resulting in a total topological charge of W = 0. The foregoing elucidations confirm that unlike figures (3b,3c) and (5d,5e), despite pivotal parameters like a, c, and $\omega$ directly swaying the count of topological charges, their total topological charge predominantly tallies to +1. Nonetheless, figures (3b,3c) and (5d,5e) stand out with divergent outcomes from other instances, marking a noteworthy observation. In figures (1a, 1c, 1e, 2a, 2c, 3a, 4a, 4c, 5a, 5c), we charted the trajectory corresponding to equation (17) across varied free parameter values. We summarized the results in Table 1
\begin{figure}[h!]
 \begin{center}
 \subfigure[]{
 \includegraphics[height=5.7cm,width=8cm]{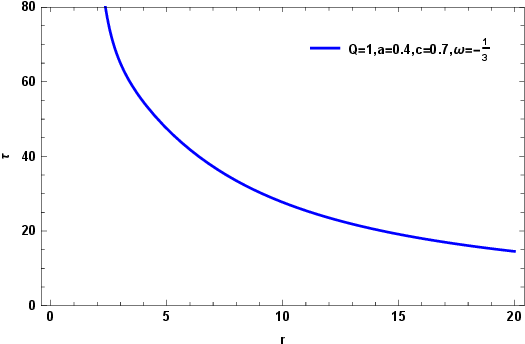}
 \label{1a}}
 \subfigure[]{
 \includegraphics[height=5.7cm,width=8cm]{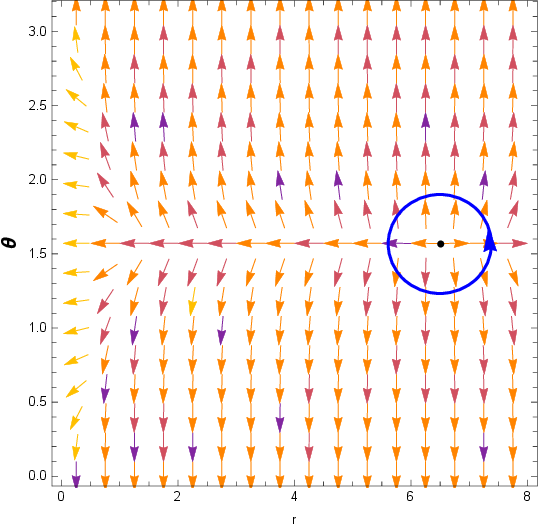}
 \label{1b}}
 \subfigure[]{
 \includegraphics[height=5.7cm,width=8cm]{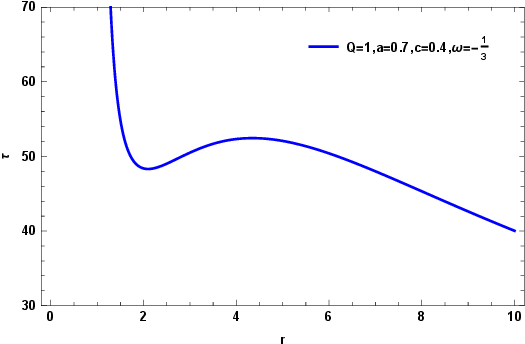}
 \label{1c}}
 \subfigure[]{
 \includegraphics[height=5.7cm,width=8cm]{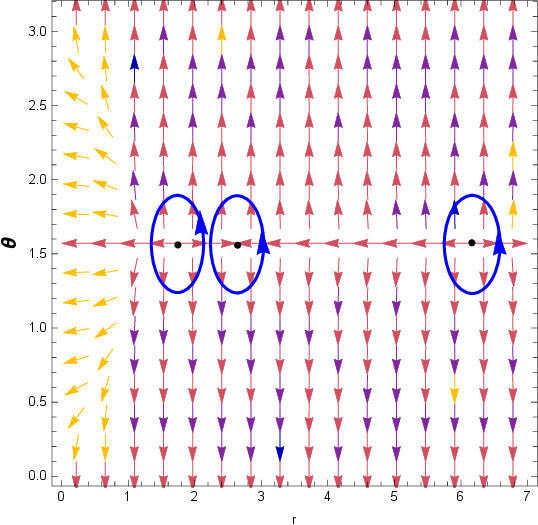}
 \label{1d}}
 \subfigure[]{
 \includegraphics[height=5.7cm,width=8cm]{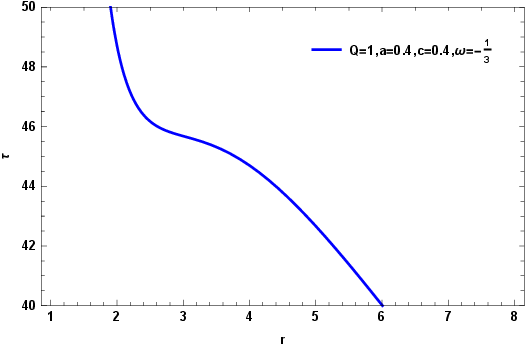}
 \label{1e}}
 \subfigure[]{
 \includegraphics[height=5.7cm,width=8cm]{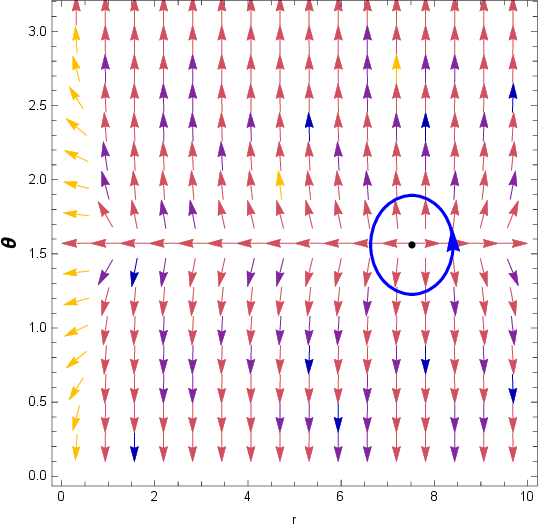}
 \label{1f}}
  \caption{\small{The plot of the curve of equation (17) with respect to (Q = 1; a = 0.4; c = 0.7) in figure (1a), (Q = 1; a =0.7; c = 0.4) in figure (1c) and (Q = 1; a = 0.4; c = 0.4) in Fig (1e) for $\omega= -1/3$. In figure (1b,1d,1f), the blue arrows represent the vector field n on a portion of the $(r-\theta)$ plane for the quantum-corrected (AdS-RN) black holes in Kiselev spacetime. The blue loops enclose the ZPs.}}
 \label{1}
 \end{center}
 \end{figure}

\begin{figure}[h!]
 \begin{center}
 \subfigure[]{
 \includegraphics[height=6.5cm,width=8cm]{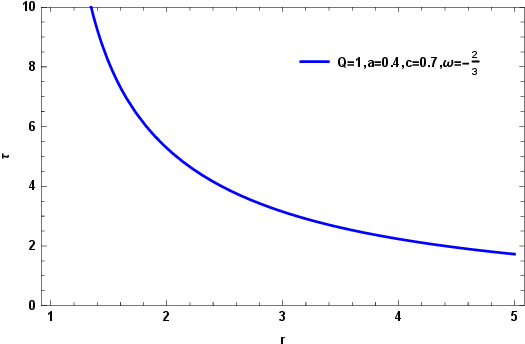}
 \label{2a}}
 \subfigure[]{
 \includegraphics[height=6.5cm,width=8cm]{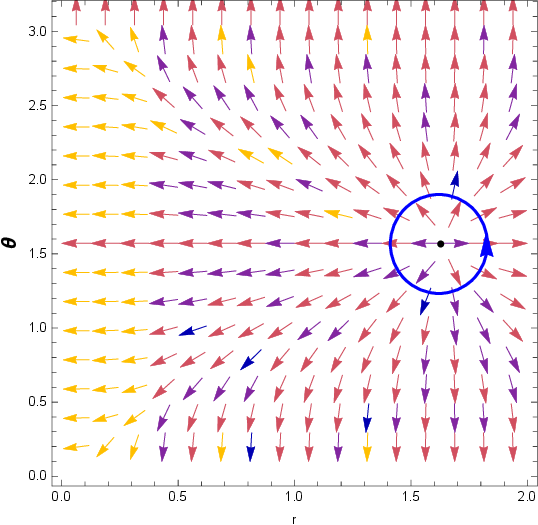}
 \label{2b}}
 \subfigure[]{
 \includegraphics[height=6.5cm,width=8cm]{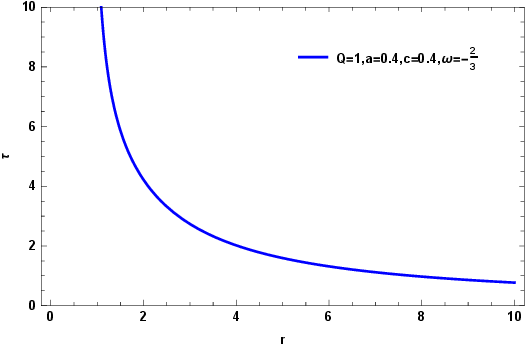}
 \label{2c}}
 \subfigure[]{
 \includegraphics[height=6.5cm,width=8cm]{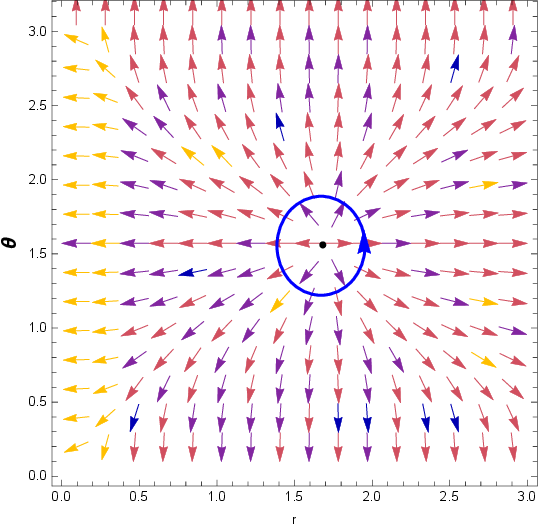}
 \label{2d}}
  \caption{\small{The plot of the curve of equation (17) with respect to (Q = 1; a = 0.4; c = 0.7) in figure (2a) and (Q = 1; a =0.4; c = 0.4) in figure (2c) for $\omega= -2/3$. In figure (2b,2d), the blue arrows represent the vector field n on a portion of the $(r-\theta)$ plane for the quantum-corrected (AdS-RN) black holes in Kiselev spacetime. The blue loops enclose the ZPs.}}
 \label{2}
 \end{center}
 \end{figure}

\begin{figure}[h!]
 \begin{center}
 \subfigure[]{
 \includegraphics[height=6.5cm,width=6cm]{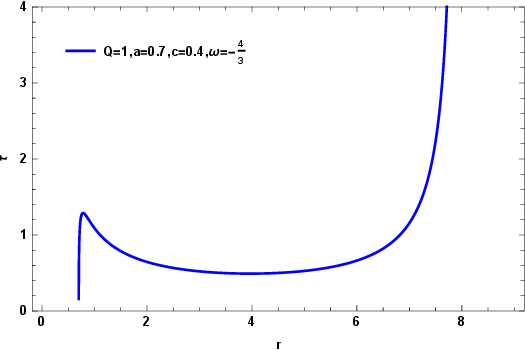}
 \label{3a}}
 \subfigure[]{
 \includegraphics[height=6.5cm,width=6cm]{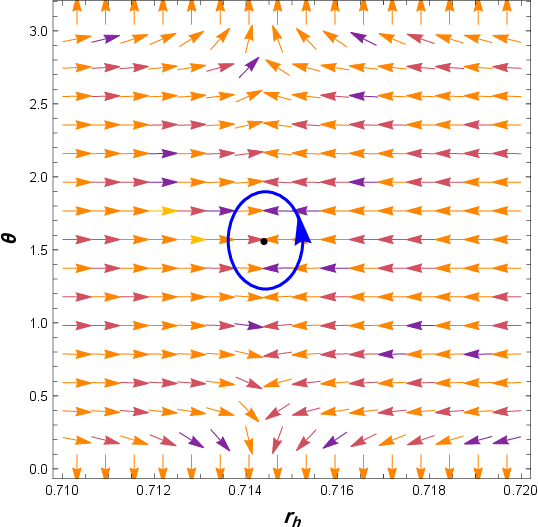}
 \label{3b}}
 \subfigure[]{
 \includegraphics[height=6.5cm,width=6cm]{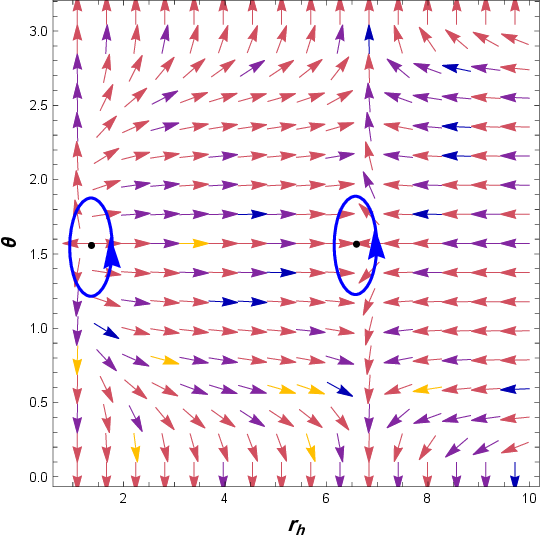}
 \label{3c}}
  \caption{\small{The plot of the curve of equation (17) with respect to (Q = 1; a = 0.7; c = 0.4) in figure (3a) for $\omega = -4/3$. In figures (3b) and (3c), the blue arrows represent the vector field n on a portion of the ($r-\theta$) plane for the quantum-corrected (AdS-RN) black holes in Kiselev spacetime. The blue loops enclose the ZPs.}}
 \label{3}
 \end{center}
 \end{figure}

\begin{figure}[h!]
 \begin{center}
 \subfigure[]{
 \includegraphics[height=6.5cm,width=8cm]{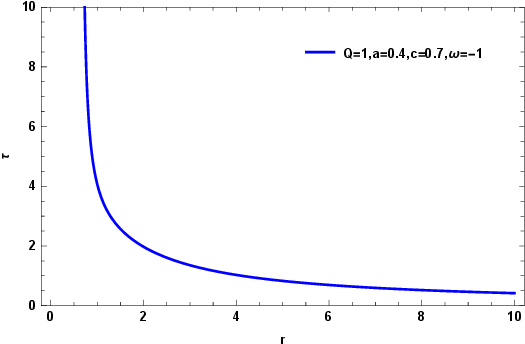}
 \label{4a}}
 \subfigure[]{
 \includegraphics[height=6.5cm,width=8cm]{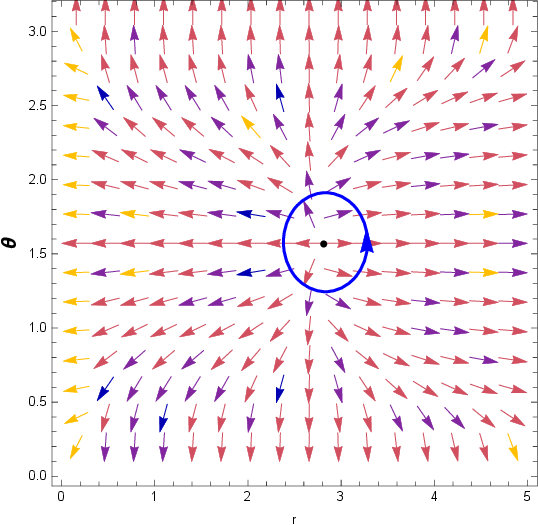}
 \label{4b}}
 \subfigure[]{
 \includegraphics[height=6.5cm,width=8cm]{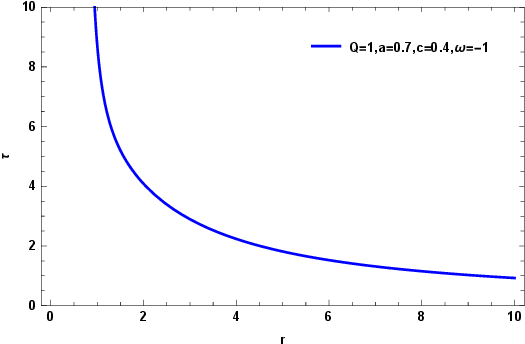}
 \label{4c}}
 \subfigure[]{
 \includegraphics[height=6.5cm,width=8cm]{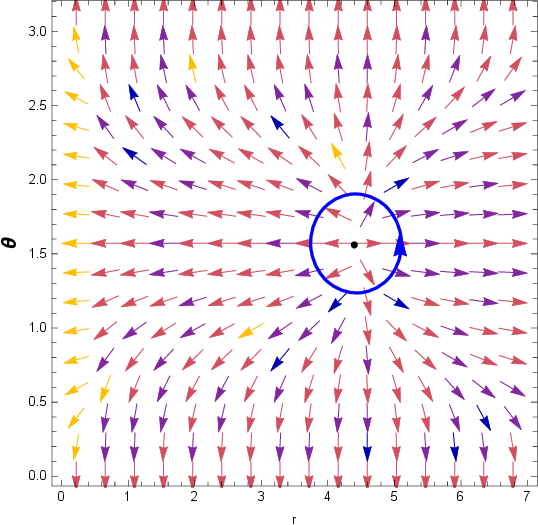}
 \label{4d}}
  \caption{\small{The plot of the curve of equation (17) with respect to (Q = 1; a = 0.4; c = 0.7) in figure (4a) and (Q = 1; a =0.7; c = 0.4) in figure (4c) for $\omega= -1$. In Fig (4b,4d), the blue arrows represent the vector field n on a portion of the ($r-\theta$) plane for the quantum-corrected (AdS-RN) black holes in Kiselev spacetime. The blue loops enclose the ZPs.}}
 \label{4}
 \end{center}
 \end{figure}

 \begin{figure}[h!]
 \begin{center}
 \subfigure[]{
 \includegraphics[height=6.5cm,width=8cm]{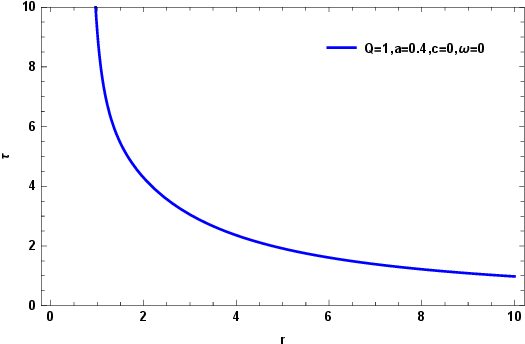}
 \label{5a}}
 \subfigure[]{
 \includegraphics[height=6.5cm,width=8cm]{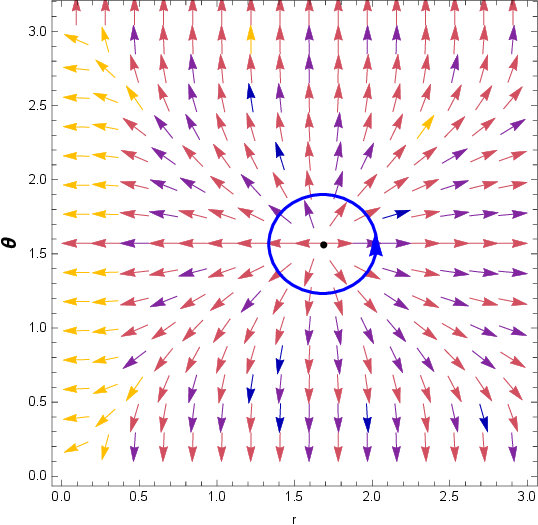}
 \label{5b}}
 \subfigure[]{
 \includegraphics[height=5.5cm,width=6cm]{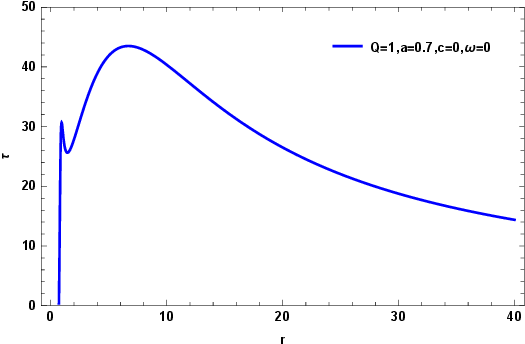}
 \label{5c}}
 \subfigure[]{
 \includegraphics[height=5.5cm,width=6cm]{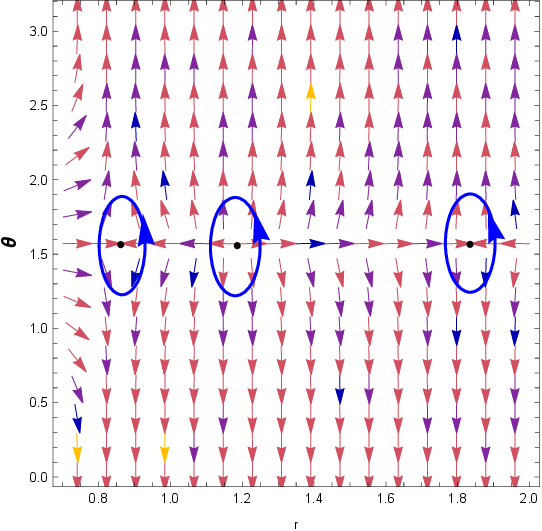}
 \label{5d}}
 \subfigure[]{
 \includegraphics[height=5.5cm,width=6cm]{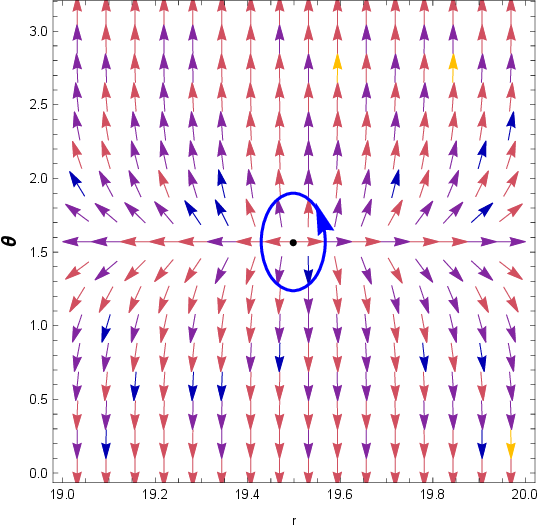}
 \label{5e}}
  \caption{\small{The plot of the curve of equation (17) with respect to (Q = 1; a = 0.4; c = 0) in figure (5a) and (Q = 1; a =0.7; c = 0) in figure (5c) for $\omega= 0$. In figure (5b,5d,5e), the blue arrows represent the vector field n on a portion of the $(r-\theta)$ plane for the quantum-corrected (AdS-RN) black holes in Kiselev spacetime. The blue loops enclose the ZPs.}}
 \label{5}
 \end{center}
 \end{figure}
 \begin{center}
\begin{table}
  \centering
\begin{tabular}{|p{6.8cm}|p{3cm}||p{3cm}|}
  \hline
   \hspace{1.3cm}Free parameters  & \hspace{1.2cm} $\widetilde{\omega}$  & \hspace{1.2cm} $W$ \\[3mm]
   \hline
    $ (Q = 1; a = 0.4; c = 0.7; \omega=-1/3)$ &\hspace{1cm} +1 & \hspace{1cm} +1 \\[3mm]
   \hline
    $ (Q = 1; a =0.7; c = 0.4; \omega=-1/3)$ & \hspace{0.7cm} +1,-1,+1 & \hspace{1cm}  +1 \\[3mm]
  \hline
   $ (Q = 1; a = 0.4; c = 0.4; \omega=-1/3)$ & \hspace{1cm}  +1 & \hspace{1cm}  +1 \\[3mm]
  \hline
  $ (Q = 1; a = 0.4; c = 0.7; \omega=-2/3)$ & \hspace{1cm} +1 & \hspace{1cm} +1  \\[3mm]
  \hline
  $(Q = 1; a =0.4; c = 0.4; \omega=-2/3)$ & \hspace{1cm} +1 & \hspace{1cm} +1 \\[3mm]
  \hline
  $(Q = 1; a =0.7; c = 0.4; \omega=-4/3)$ & \hspace{0.8cm} -1,+1,-1 & \hspace{1.2cm} -1 \\[3mm]
  \hline
  $(Q = 1; a = 0.4; c = 0.7; \omega=-1)$ & \hspace{1cm} +1 & \hspace{1cm} +1 \\[3mm]
  \hline
  $(Q = 1; a =0.7; c = 0.4; \omega=-1)$ & \hspace{1cm} +1 & \hspace{1cm} +1 \\[3mm]
  \hline
  $(Q = 1; a = 0.4; c = 0; \omega=0)$ & \hspace{1cm} +1 & \hspace{1cm} +1 \\[3mm]
  \hline
  $(Q = 1; a =0.7; c = 0; \omega=0)$ & \hspace{0.7cm}  -1,+1,-1,+1 & \hspace{1cm} 0 \\[3mm]
  \hline
\end{tabular}
\caption{Summary of the results}\label{1}
\end{table}
 \end{center}

\section{Conclusions}
In this article, we explore the thermodynamic topology of quantum-corrected AdS-RN black holes in the Kiselev spacetime. Utilizing the advanced generalized off-shell Helmholtz free energy methodology, we perform a detailed calculation of the thermodynamic topology for this black hole. Additionally, we rigorously study their topological classifications. Our research uncovers that integrating quantum correction terms within Kiselev spacetime plays a pivotal role in the topological classifications of black holes. This leads us to uncover groundbreaking insights into their intrinsic characteristics and behaviors, potentially reshaping our understanding of black hole physics. In conclusion, our investigation into the thermodynamic topology of a quantum-corrected charged AdS black hole within Kiselev spacetime has yielded significant insights. The study meticulously analyzed the zero points across various figures, which signify the presence of topological charges. These topological charges are linked to the winding number and are encapsulated within the blue contour loops at specific coordinates $(r, \theta)$. The parameter $\omega$ plays a pivotal role in dictating the topological charges, with values ranging from $\omega=-1/3, -2/3, -4/3, -1$ and 0, influencing the topological charge. The free parameters chosen for this study were Q = 1 and (a and c = 0.4; 0.7).  Notably, figures (5d,5e) present an exception, depicting four topological charges that result in a total topological charge of zero, and figures (3b,3c) have three topological charges that result in a total topological charge of $-1$ deviating from the otherwise consistent total topological charge $W=+1$ observed in other figures. This anomaly occurs at $\omega=-4/3$ and $\omega=0$. Furthermore, certain figures display three topological charges, culminating in a total topological charge $W=+1$, highlighting the influence of the parameters a, c, and $\omega$ on the topological charge count. Despite the variations introduced by these parameters, the study concludes that the total topological charge predominantly remains at +1, affirming the thermodynamic stability of the black hole, except for the figures (3b,3c) and (5d,5e).

\end{document}